\documentclass[aps,prl,twocolumn,superscriptaddress]{revtex4-1}

\usepackage{verbatim}
\usepackage{amsmath}
\usepackage[utf8]{inputenc}
\usepackage[pdftex]{graphicx}
\usepackage[thinspace]{SIunits}
\usepackage[colorlinks=true,urlcolor=blue,linkcolor=blue,citecolor=blue]{hyperref}
\usepackage{xcolor}
\usepackage[T1]{fontenc}
\usepackage{placeins}  
\usepackage{nicefrac}
\usepackage{braket}

\begin{document}

\title{A few-emitter solid-state multi-exciton laser}
%
%
%
\author{S.~Lichtmannecker}
\affiliation{ 
Walter Schottky Institut and Physik Department, Technische Universit\"at M\"unchen, Am Coulombwall 4, 85748 Garching, Germany}
\author{M.~Florian}
\affiliation{
Institut für theoretische Physik,
Universit\"at Bremen, Otto-Hahn-Allee 1, 28359 Bremen}
\author{T.~Reichert}
\affiliation{
Walter Schottky Institut and Physik Department, 
Technische Universit\"at M\"unchen, Am Coulombwall 4, 85748 Garching, Germany}
\author{M.~Blauth}
\affiliation{
Walter Schottky Institut and Physik Department, 
Technische Universit\"at M\"unchen, Am Coulombwall 4, 85748 Garching, Germany}
\author{M.~Bichler}
\affiliation{
Walter Schottky Institut and Physik Department, 
Technische Universit\"at M\"unchen, Am Coulombwall 4, 85748 Garching, Germany}
\author{F.~Jahnke}
\affiliation{
Institut für theoretische Physik,
Universit\"at Bremen, Otto-Hahn-Allee 1, 28359 Bremen}
\author{J.~J.~Finley}\email{jonathan.finley@wsi.tum.de}
\affiliation{
Walter Schottky Institut and Physik Department, 
Technische Universit\"at M\"unchen, Am Coulombwall 4, 85748 Garching, Germany}
\author{C.~Gies}
\affiliation{
Institut für theoretische Physik,
Universit\"at Bremen, Otto-Hahn-Allee 1, 28359 Bremen}
\author{M.~Kaniber}\email{michael.kaniber@wsi.tum.de}
\affiliation{
Walter Schottky Institut and Physik Department, 
Technische Universit\"at M\"unchen, Am Coulombwall 4, 85748 Garching, Germany}
%

\date{\today}

\begin{abstract}
We report a combined experimental and theoretical study of non-conventional lasing from higher multi-exciton states of a few quantum dot-photonic crystal nanocavity. We show that the photon output is fed from saturable quantum emitters rather than a non-saturable background despite being rather insensitive to the spectral position of the mode. Although the exciton transitions of each quantum dot are detuned by up to $160$ cavity linewidths, we observe that strong excitation populates a multitude of closely spaced multi-exciton states, which partly overlap spectrally with the mode. The limited number of emitters is confirmed by a complete saturation of the mode intensity at strong pumping, providing sufficient gain to reach stimulated emission, whilst being accompanied by a distinct lasing threshold. Detailed second-order photon-correlation measurements unambiguously identify the transition to lasing for strong pumping and, most remarkably, reveal super-thermal photon bunching with $g^{(2)}(0)>2$ below lasing threshold. Based on our microscopic theory, a pump-rate dependent $\beta$-factor $\beta(P)$ is needed to describe the nanolaser and  account for the interplay of multi-exciton transitions in the few-emitter gain medium. Moreover, we theoretically predict that the super-thermal bunching is related to dipole-anticorrelated multi-exciton recombination channels via sub- and super-radiant coupling below and above lasing threshold, respectively. Our results provide new insights into the microscopic light-matter-coupling of spatially separated emitters coupled to a common cavity mode and, thus, provides a complete understanding of stimulated emission in nanolasers with discrete emitters.
\end{abstract}

\maketitle
%

%
%
Self-assembled semiconductor quantum dots (QDs) exhibit richer discrete
energy level structures compared to atoms, due to their mesoscopic size, shape
and dielectric surroundings \cite{bimberg1999quantum}. When embedded within nanostructured 
photonic cavities, QDs allow for the investigation of cavity quantum
electrodynamics phenomena in the solid state and provide strong potential for
photon mediated quantum information technologies in a uniquely scalable
architecture \cite{o2009photonic}. Photonic crystal (PhC)
\cite{joannopoulos1997photonic} nanostructures in particular provide strong
optical confinement in high quality (Q) factor and small mode volume
($V_\mathrm{mode}$) cavities \cite{noda2007spontaneous}, making them suitable to
explore the miniaturization limit of lasing where the gain medium consists only
of a few solid-state quantum emitters within a single mode cavity. Recent experiments perfromed on PhC cavities have
demonstrated stimulated emission and lasing
\cite{hendrickson2005quantum,reitzenstein2006lasing,strauf2006self,altug2006ultrafast}.
Moreover, the rich multi-exciton structure provided by the QDs has been shown to play a
significant role in far-off resonant cavity feeding and photon bunching from
PhC cavities \cite{winger2009explanation,laucht2010temporal}. Similar results of
photon bunching and a clear transition to coherent emission and lasing operation
were demonstrated in micropillar cavities subject to optical excitation or electrical injection
\cite{ulrich2007photon,schneider2013electrically}.

In this letter, we observe a transition from spontaneous to non-conventional lasting from an L3 PhC nanocavity loaded with a few (N
$ \sim 4$) spectrally detuned QDs. The intensity of the cavity mode emission as a
function of the excitation level exhibits a weak threshold, indicative of a transiton to lasing for a level very close to the saturation power density of nearby QDs
$P_\mathrm{sat}^\mathrm{QD} \sim \unit{0.14 \pm
0.1}{\kilo\watt\per\centi\meter\squared}$. Moreover, the input-output characteristic of the device
shows a complete saturation for excitation power densities of
$P_{sat}^{cav}>\unit{4.7\pm 0.4}{\kilo\watt\per\centi\meter\squared}$,
demonstrating that the cavity mode emission is fed from saturable emission from a few QDs, rather than a broadband non-saturable background \cite{strauf2006self}.
The transition to lasing is identified via second-order photon-correlation spectroscopy performed on the detuned cavity-mode emission that shows
super-thermal photon bunching at zero time delay with \smash{$g^{(2)}(0)>2$} at
low power densities, and an unambiguous transition to coherent light emission
with \smash{$g^{(2)}(0)=1$} at elevated excitation. Remarkably, the experimental results
are shown to be insensitive to the mean detuning between the QDs and the cavity. Insights into the
underlying physical mechanisms responsible for lasing are gained from theoretical simulations that take into
account multi-exciton states of four QDs spatially located in the cavity region,
and their non-perturbative interaction with the photons in the laser mode. On
the basis of the theoretical model, we introduce a pump-rate dependent $\beta$-factor $\beta(P)$ that
characterizes the spontaneous emission coupling into the cavity mode. We show
that the interplay of multi-exciton transitions in the few-emitter gain medium
gives rise to a strong pump-density dependence of $\beta(P)$ . As such, we show that the system is not
well described in terms of the constant $\beta$-factor that is used in conventional
laser theories. Radiative coupling between spatially separated emitters mutually coupled to the stronlgy confined cavity field has recently
been predicted to lead to sub- and superradiant effects in nanolasers
\cite{leymann2015sub,mascarenhas2013cooperativity}. From our theoretical
analysis, the super-thermal bunching is identified to arise from dipole-anticorrelated
multi-excitonic emission channels that emit subradiant light
below threshold. The transition from subradiant emission to superradiant lasing
is also reflected in the pump-dependent factor $\beta(P)$.
Our work provides new insights tothe lasing mechanism, spontaneous emission coupling 
factor and radiative QD-QD coupling in few-dot nanocavity lasers.
%
%
%
%
\begin{figure}
\includegraphics{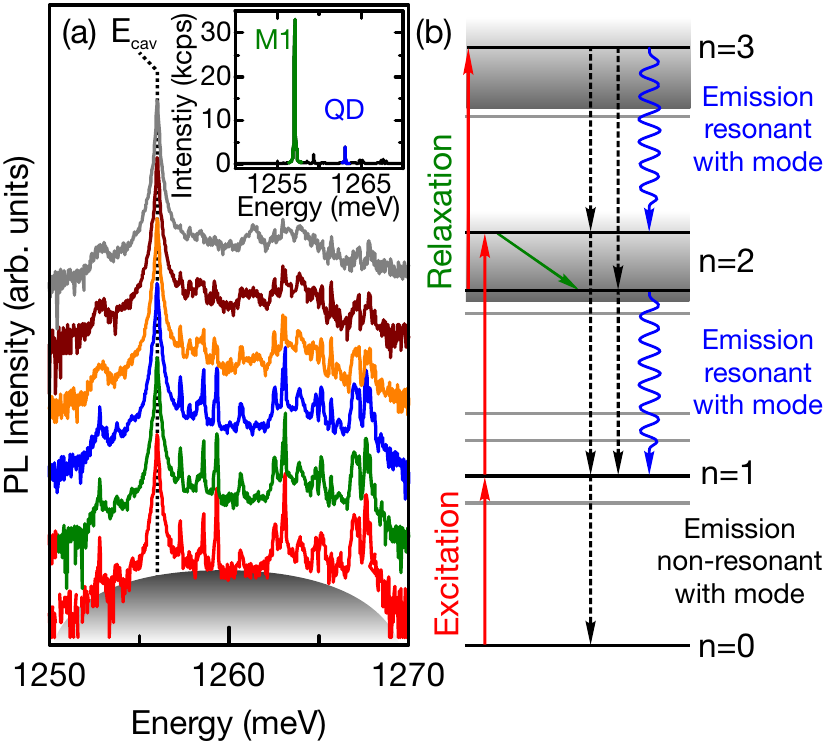}
    \renewcommand{\figurename}{Figure}
\caption{\label{figure_1}%
(a) Emission spectra from the cavity increasing cw excitation power
density from $\unit{0.14}{\kilo\watt\per\centi\meter\squared}$ to
$\unit{5.9}{\kilo\watt\per\centi\meter\squared}$. The spectra are plotted on a
logarithmic scale with offset to each other for clarity. The cavity mode is
labeled $E_\mathrm{cav}$. The inset shows a linear spectrum of the system for an
excitation power density of $P^\mathrm{QD}_\mathrm{sat}$.
(b) Schematic level schema of the manifolds of multi-excitonic emission channels
with $n=0\ldots 3$ excitations. The arrows indicate the physical excitation,
relaxation, and recombination processes that are included in the theoretical
modeling, as explained in the text.
}
\end{figure}

To characterize our system, we show in figure~\ref{figure_1} (a) 
$\unit{}{\micro}$-PL spectra on a stacked logarithmic scale recorded with cavity mode
resonant excitation via the third order cavity mode
\cite{chalcraft2007mode,kaniber2008tunable} using an excitation power density
increasing from $\unit{0.14}{\kilo\watt\per\centi\meter\squared}$ (bottom spectrum) to
$\unit{5.9}{\kilo\watt\per\centi\meter\squared}$ (top spectrum). The spectra show the cavity
mode emission at $E_\mathrm{cav}=\unit{1257.1}{\milli\electronvolt}$ and the emission 
of a few (N $\sim4$) QDs located in the PhC nanocavity. The emission of these
QDs evolves into a broadband emission attributed to multi-exciton transitions
for elevated excitation intensities, as highlighted by the gray shaded region
\cite{winger2009explanation}.  The inset shows a spectrum in a linear
representation for an excitation power density of
$\unit{0.14}{\kilo\watt\per\centi\meter\squared}$ close to the saturation
power density $P^\mathrm{QD}_\mathrm{sat}$ of the single QD emitting at
$E_\mathrm{QD}=\unit{1263.1}{\milli\electronvolt}$, highlighted in blue. Fitting
the fundamental cavity mode M1 (highlighted in green) with a Lorentzian line yields a full
width at half maximum of $\Delta E = \unit{104 \pm
2}{\micro\electronvolt}$ corresponding to a moderate mode $Q= E_\mathrm{cav}/\Delta E \approx 12000$. Measuring the autocorrelation function of the
QD emission at $E_{QD}$ reveals the single-photon character of the emission, and the
cross correlation measurement of the QD and the M1 cavity mode (shown in the Supplementary Material)
proves non-resonant coupling through antibunched emission despite the
large initial energy detuning of $\Delta_\mathrm{QD-M1}= \unit{6}{\milli\electronvolt}$
\cite{hennessy2007quantum,press2007photon,kaniber2008investigation,michler2000quantum,santori2001triggered}.

On the basis of this characterization, we consider a theoretical model that
accounts for the key elements of the experiment, namely the interplay of multi-excitonic 
emission channels from several different QD emitters, and their light-matter 
interaction with photons in the cavity mode. We assume that each QD
possesses a multitude of many-particle states represented by different number
$n$ of excitations in the form of e-h-pairs and their distribution over the
available many-particle states \cite{laucht2010temporal}. For the $n=1$ to $n=0$
transition only well-separated emission lines exist, whereas for $n=2$ to $n=1$
and $n=3$ to $n=2$ a large number of possible transitions result in a broad
range of closely spaced lines resembling a quasi-continuous spectrum
\cite{winger2009explanation}, as illustrated in figure~\ref{figure_1} (b). To
model the emission characteristics, we solve the von Neumann-Lindblad equation
for the density matrix of the coupled carrier-photon system with
non-perturbative Jaynes-Cummings interaction. Due to the large state space,
calculations can only be performed for a subset of the QD many-particle
configurations. For each QD, we choose two multi-exciton configurations whose
recombination is resonant with the cavity mode, whereas the ground state exciton
transition is strongly detuned. This allows us to study the interplay of QD
many-particle states and the resulting coupling to the cavity mode within a wide
spectral window. Excitation and relaxation processes indicated in
figure~\ref{figure_1}(b) are described by Lindblad terms. Details of the
microscopic model are found in the supplementary material.

%
\begin{figure*}
\includegraphics{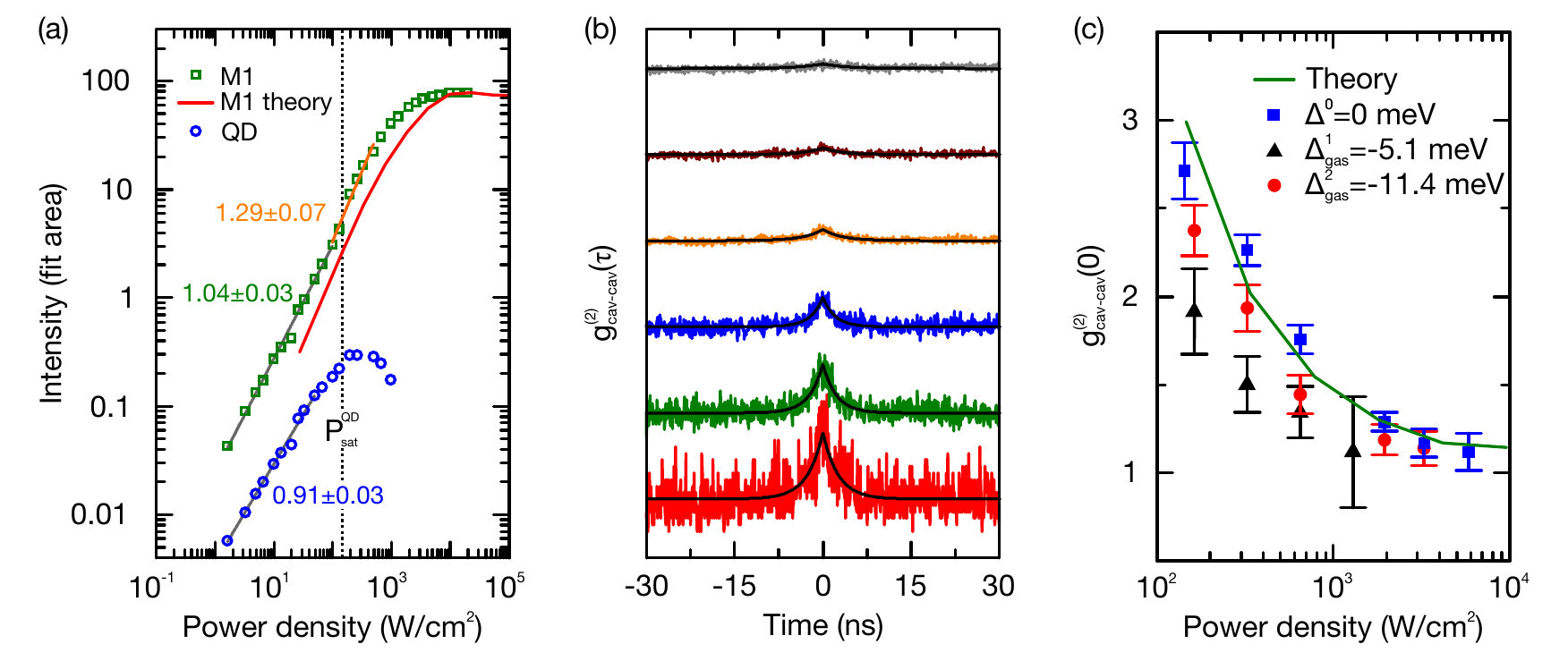}
    \renewcommand{\figurename}{Figure}
\caption{\label{figure_2}%
(a) Integrated intensity of the cavity mode (green) and the QD (blue) as a
function of excitation power density. Black solid lines represent power-law fits
to the emission data. The solid red line depicts the intensity of the cavity mode
emission calculated from theory. To connect the theoretical pump rate with the
experimental power density, the red curve has been shifted along the power axis to ensure that the calculated exciton saturation coincides with
$P_\mathrm{sat}^\mathrm{QD}$.
(b) Auto-correlation measurements of the cavity mode for the accordingly color
coded spectra  presented in figure~\ref{figure_1} (a) ($\Delta^0$ in figure
\ref{figure_2} (c)). Solid black lines are fits to the data.
(c) Second oder correlation \smash{$g_\mathrm{cav-cav}^{(2)}(0)$} as function of
the excitation power density. The colors represent three different cavity mode
energies, blue $\Delta^0 = \unit{0}{\milli\electronvolt} \equiv E_\mathrm{cav}$,
black $\Delta^1_\mathrm{gas} = \unit{-5.1}{\milli\electronvolt}$ and red
$\Delta^2_\mathrm{gas} = \unit{-11.4}{\milli\electronvolt}$. The solid green
line is the numerical result  \smash{$g^{(2)}(0)$} from theory.
}
\end{figure*}

\begin{figure}
\includegraphics{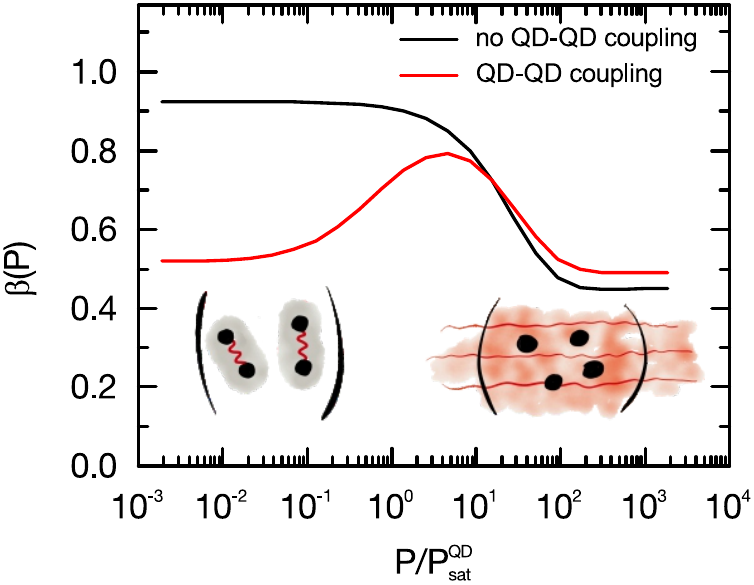}
    \renewcommand{\figurename}{Figure}
\caption{
(Red curve) pump-rate dependent $\beta$-factor obtained from the theoretical model with parameters applicable to experimentally studied system. 
(Black curve) calculation suppressing radiative coupling effects between emitters that are
responsible for sub- and superradiant effects. Comparing
both curves reveals that radiative coupling effects lead to a strong inhibition
of spontaneous emission at low exitation (subradiance) and a slight enhancement
of spontaneous emission above the laser threshold
(superradiance).}
\label{figure_3}
\end{figure}

In figure~\ref{figure_2} (a) we present the integrated (green symbols) and
calculated (red line) PL intensity of the cavity as a function of excitation
power density, as well as the QD emission (blue symbols). Fitting a power law $I
= A \cdot P^m$ to the intensity $I$ of the QD reveals a linear behaviour
with an exponent $m^\mathrm{QD} = 0.91\pm0.03$  as shown by the black line, indicating
single excitonic character of the emission line \cite{finley2001charged}.
Moreover, the QD emission saturates at an excitation power density of
$P_\mathrm{sat}^\mathrm{QD}=\unit{0.14 \pm 0.1}{\kilo\watt\per\centi\meter\squared}$ as
highlighted by the dotted black line in figure~\ref{figure_2} (a). For the cavity mode emission we observe a similar behavior for excitation power
densities $P<P_\mathrm{sat}^\mathrm{QD}$, reflected by an exponent $m^\mathrm{cav}_1=1.06 \pm 002$.
For excitation power densities $P_\mathrm{sat}^\mathrm{QD}<P<P_\mathrm{sat}^\mathrm{cav}$ we observe a
slight superlinear increase, giving rise to an exponent of
$m^\mathrm{cav}_2=1.22\pm0.07$, highlighted in orange in figure~\ref{figure_2} (a). The
increase in the exponent of the cavity intensity power dependence appears simultaneously
with the saturation of the QD, when multi-exciton states become increasingly populated with
significant probability. For even higher excitation power densities
$P>P_\mathrm{sat}^\mathrm{cav}=\unit{4.7 \pm 0.4}{\kilo\watt\per\centi\meter\squared}$ a
complete saturation of the mode emission is observed. This finding unambiguously confirms the absence of non-saturable background 
contributions and reflects the limited gain the few QDs are able to provide. The calculated input-output-characteristic (red curve) reproduces
the main features that are seen in the experimental data, namely a nearly
thresholdless increase and full saturation of the output.  The slight kink in the input-ouput
characteristic of the mode emission in combination with the observed saturation
behavior gives rise to a moderate s-shape curve typical for nanolasers
\cite{bjork1992linewidth} with ultra-low thresholds
\cite{strauf2006self,reitzenstein2006lasing}. Procedures from rate equations for
conventional lasers might suggest to estimate a $\beta$-factor from this kink,
see e.g. Ref. \cite{strauf2006self,michler2000quantum}, however, we will demonstrate that the
underlying mechanism of multi-exciton lasing requires an entirely different approach, as we explain below in the
context of figure~\ref{figure_3}.

It is well recognized that photon autocorrelation measurements provide a clear
indication for lasing in the absence of a visible threshold in the input-output
characteristics. In order to support our previous hypothesis of a low-threshold,
few-QD nanolaser, we present detailed investigations of \smash{$g^{(2)}_\mathrm{cav-
cav}(\tau)$} for the cavity mode emission. For each of the color-coded
excitation power densities shown in figure~\ref{figure_1} (a) we performed
second-order photon correlation measurements of the cavity mode \smash{$g^{(2)}_\mathrm
{cav-cav}(\tau)$} which are presented in figure~\ref{figure_2} (b).
Fitting the data with \mbox{\smash{$g^{(2)}(\tau)=1+A\cdot
\exp(-2|\tau|/t_0)$}} (shown by the black lines)
\cite{loudon2000quantum} enables us to extract the zero-time-delay values of
\smash{$g_\mathrm{cav-cav}^{(2)}(0)$} shown by the blue symbols in
figure~\ref{figure_2} (c). Upon increasing the excitation level from
$\unit{0.14}{\kilo\watt\per\centi\meter\squared}$ to
$\unit{5.9}{\kilo\watt\per\centi\meter\squared}$ we demonstrate a clear
transition  from the spontaneous-emission regime with \smash{$g^{(2)}_\mathrm{cav-
cav}(0)>>1$} to lasing with \smash{$g^{(2)}_\mathrm{cav-cav}(0)=1$}
\cite{ulrich2007photon}. The photon autocorrelation function is also readily
available within the density-matrix formalism  \cite{florian2013phonon}, and the
theoretical model predicts the same qualitative and quantitative behavior, as
shown by the green line in figure~\ref{figure_2} (c).

Interestingly, in the low-excitation regime super-thermal values up to
\smash{$g^{(2)}(0)=2.71\pm0.16$} are observed in figure~\ref{figure_2} (c) both
in theory and experiment. They exceed the values reported in previous
experimental studies \cite{strauf2006self,winger2009explanation} by a factor of
2, and even the theoretical limit of 2 for thermal light.
The enhanced probability of two- and multiple-photon emission events reflected by
\smash{$g^{(2)}(0)>2$}  is attributed to the presence of competing resonant
emission channels for each QD emitter, allowing for the simultaneous emission of
photons into the mode. Moreover, in the numerical calculation we observe the
presence of strong radiative coupling effects (see the discussion of
figure~\ref{figure_3}). Such effects have been predicted to leave a
super-thermal finger print in cw-excited nanolasers at low-excitation powers
\cite{leymann2015sub,mascarenhas2013cooperativity,auffeves2011few}. We interpret
the experimental observation of the super-thermal bunching as proof for a new
regime of spontaneous emission with radiatively enhanced correlations between
distant emitters inside the nanocavity.

To shed more light on the efficiency of non-resonant coupling, we have studied
the influence of cavity emitter detuning $\Delta$ on the cavity-mode correlation
function $g_\mathrm{cav-cav}^{(2)}(\tau)$ by red-shifting the cavity mode energy by locally freezing 
inert nitrogen into the PhC \cite{mosor2005scanning,kaniber2008tunable}.
Figure~\ref{figure_2} (c) shows the experimentally determined values of
\smash{$g_\mathrm{cav-cav}^{(2)}(0)$} as a function of excitation power density for
three different cases with dot-cavity detunings, $\Delta^0 =
\unit{0}{\milli\electronvolt}$ (blue squares), $\Delta_\mathrm{gas}^1 =
\unit{-5.1}{\milli\electronvolt}$ (black triangles) and $\Delta_\mathrm{gas}^2 =
\unit{-11.4}{\milli\electronvolt}$ (red circles). Similar behavior of \smash{$g_\mathrm
{cav-cav}^{(2)}(0)$} is observed for all detunings, namely a clear transition from
spontaneous emission to lasing.  Coherent emission is clearly reached at
$P=\unit{10^3}{\watt\per\centi\meter\squared}$ 
for all cases before the onset of cavity mode saturation. Furthermore, we
observe that the efficiency of the non-resonant coupling makes nanolaser
operation rather robust for total spectral cavity-emitter detunings up to $\Delta \sim
\unit{17}{\milli\electronvolt}$. Thus, the absolute energies of QD-transitions and cavity mode are of limited importance for the operation of the nanolaser. This finding further motivates
the simplification of considering only resonant multi-excition transitions in
the microscopic modeling.

We now turn to the characterization of spontaneous emission coupling in the
presence of multi-exciton lasing. The $\beta$-factor quantifies the fraction of
the total spontaneous emission that is directed into the laser mode. From a rate
equation model it is found that $\beta$ is solely determined by the relation of
the rates associated with emission into the laser mode $\gamma$ and into
non-lasing modes or other loss channels $\gamma_\mathrm{nl}$, i.e.,
$\beta=\gamma/(\gamma+\gamma_\mathrm{nl})$. In the rate equations, the
rates are understood to be ensemble averages of two-level systems. However, if the gain
material consists of only few solid-state emitters,  multi-exciton transitions
from different emitters may tune in and out of resonance with the cavity mode as pumping is varied.
Thus, the coupling efficiency to the cavity mode varies for each of these emission
channels, which must be accounted for when formulating the $\beta$-factor for few-QD nanolasers. 
To obtain a quantity that reflects this behavior, we calculate for each pump rate
the averaged spontaneous emission rate of the QD ensemble from the spontaneous
emission contribution $\Gamma$ to the photon-assisted polarization
\cite{leymann2015sub}, which is given by
\begin{align}\label{eq:gamma}
    \begin{split}
    \Gamma = \sum_{i,j}&\left[\sum_{\alpha=1}^{N_{\mathrm{QD}}} R_i \Braket{\left(D^{l}_{\alpha,i}\right)^\dagger D^{l}_{\alpha,j}}\delta_{i,j} \right.\\
    &+\left.\sum_{\alpha\neq\beta}^{N_{\mathrm{QD}}}  R_i \Braket{D^{l}_{\alpha,i}\left(D^{l}_{\beta,j}\right)^\dagger}\right].
    \end{split}
\end{align}

Here, the operator $D^{l}_{\alpha,i}$ describes an allowed (bright) dipole transition
between multi-exciton states, e.g. from exciton to ground state, in QD
$\alpha$, with the initial state $\ket{i_\alpha}$, and the corresponding recombination rate $R_i$.
The quantum-mechanical average is taken with respect to the steady-state density
operator. The first sum in Eq.~\eqref{eq:gamma} includes all bright transitions,
while the sums in the brackets address the QD emitters in the gain medium. We
distinguish two contributions: The first term in the brackets is the sum of the
spontaneous emission contributions from all emitters \emph{independently}, while
the second sum is the contribution of \emph{dipole-correlated transitions in
different emitters}, which arises due to radiative coupling. In
figure~\ref{figure_3} we show the pump-rate dependent factor $\beta(P)$
calculated from this rate $\Gamma$ including (red curve) and excluding (black curve) radiative
emitter coupling, and under consideration of the loss-rates as given by the
Lindblad terms. For further details of the theoretical description, we refer to the Supplementary Material. Without
radiative coupling effects (black curve), a transition is seen from $\beta(P)>
90\%$ to $\beta(P)\approx 50\%$ as the system switches between multi-excitonic
emission channels from different manifolds, as schematically depicted in figure~\ref{figure_1}(b). These
limiting values at low and high excitation reflect the conventional constant
$\beta$-factor associated with these transitions. This behavior is drastically
changed due to radiative coupling effects (red curve), which cause a strong
inhibition of the spontaneous emission rate below threshold, resulting in
$\beta(P) = 50\%$ instead of $90\%$ at low excitation. Spontaneous emission
inhibition has been identified previously as subradiance due to an anti-correlation 
of dipoles in different emitters~\cite{leymann2015sub}. In Eq.~\ref{eq:gamma} it results
from a negative contribution from the second term. At excitation powers $\gtrsim
20 P_\mathrm{sat}^{QD}$, the sign changes and spontaneous emission is enhanced
due to \emph{super}radiant coupling between emitters. 

While we can directly quantify the effect of sub- and superradiant inter-emitter coupling in the spontaneous emission rate, it is not straightforward to switch off inter-emitter coupling in the numerical results shown in figure.~\ref{figure_2}, and even less so in the experiment. However, the comparison in figure~\ref{figure_3}, together with the super-thermal bunching observed at weak exciation as shown in figure~\ref{figure_2}(c), provides a strong account for the presence of radiative coupling in the QD nanocavity system that is supported by predictions made in Ref.~\cite{leymann2015sub}. 

Finally, we point out
that a kink in the input-output characteristics can be misinterpreted in terms
of a constant $\beta$-factor, but may in fact result from transitions between
multi-exciton states of various emitters tuning in and out of resonance at
various excitation powers.

In summary, we presented new insight into the extraordinary operational regime of
a few ($\sim 4$) solid-state emitter PhC nanolaser. We have
observed super-thermal bunching of the emission and explained it on the basis of
a microscopic theory to arise from the simultaneous presence of different
resonant emission channels and their radiative coupling. We have further
demonstrated that a conventional single-valued $\beta$-factor cannot
characterize the interplay of multi-exciton lasing in few-emitter QD lasers. The
newly introduced factor $\beta(P)$ is pump-rate dependent and strongly
determined by radiative coupling effects. In combination with the super-thermal
bunching, it gives strong account for the presence of radiative coupling effects
in the form of sub- and superradiance in the nanolaser system. At elevated
excitation powers, lasing is demonstrated, while full saturation of the emissions
at the highest excitation powers proves that only a few saturable multi-exciton
states contribute to the excitation, and that continuum states from the wetting
layer are not necessary for lasing operation in our device.


\paragraph{Acknowledgements:} We thank F.P. Laussy and E. del Valle for highly
fruitful discussions and gratefully acknowledge financial support from the
DFG via SFB 631, JA 619/10-3, JA 619/13-1 and GI 1121/1-1, the German Excellence Initiative via NIM, as well as from
the BMBF via QuaHL-Rep and Q.com. 

\FloatBarrier
\bibliographystyle{apsrev}
\bibliography{Bibliography_BunchingCavity}
 
\end{document}


\title{Supplementary Material: A few-emitter solid-state multi-exciton laser}
%
%
%
\author{S.~Lichtmannecker}
\affiliation{
Walter Schottky Institut and Physik Department, 
Technische Universit\"at M\"unchen, Am Coulombwall 4, 85748 Garching, Germany}
\author{M.~Florian}
\affiliation{
Institut für theoretische Physik,
Universit\"at Bremen, Otto-Hahn-Allee 1, 28359 Bremen}
\author{T.~Reichert}
\affiliation{
Walter Schottky Institut and Physik Department, 
Technische Universit\"at M\"unchen, Am Coulombwall 4, 85748 Garching, Germany}
\author{M.~Blauth}
\affiliation{
Walter Schottky Institut and Physik Department, 
Technische Universit\"at M\"unchen, Am Coulombwall 4, 85748 Garching, Germany}
\author{M.~Bichler}
\affiliation{
Walter Schottky Institut and Physik Department, 
Technische Universit\"at M\"unchen, Am Coulombwall 4, 85748 Garching, Germany}
\author{F.~Jahnke}
\affiliation{
Institut für theoretische Physik,
Universit\"at Bremen, Otto-Hahn-Allee 1, 28359 Bremen}
\author{J.~J.~Finley}\email{jonathan.finley@wsi.tum.de}
\affiliation{
Walter Schottky Institut and Physik Department, 
Technische Universit\"at M\"unchen, Am Coulombwall 4, 85748 Garching, Germany}
\author{C.~Gies}
\affiliation{
Institut für theoretische Physik,
Universit\"at Bremen, Otto-Hahn-Allee 1, 28359 Bremen}
\author{M.~Kaniber}\email{michael.kaniber@wsi.tum.de}
\affiliation{
Walter Schottky Institut and Physik Department, 
Technische Universit\"at M\"unchen, Am Coulombwall 4, 85748 Garching, Germany}
%

\date{\today}

\maketitle

\section{Fabrication and exerimental details} 
The sample investigated was grown using molecular beam epitaxy on a
$\unit{350}{\micro\metre}$ thick [100] GaAs wafer. A 300 nm GaAs buffer layer
was grown, followed by an $\unit{800}{\nano\metre}$ thick sacrificial layer of
Al$_\mathrm{0.8}$Ga$_\mathrm{0.2}$As and an $\unit{150}{\nano\metre}$ thick nominally undoped
GaAs slab that contained a single layer of nominally In$_{0.5}$Ga$_{0.5}$As  QDs at its
midpoint. The growth conditions used for the QD layer yielded an areal  density
$\rho_D$ \mbox{$\sim 20$ $\mu$m$^{-2}$}, emitting over the spectral range of
$\unit{1270-1400}{\milli\electronvolt}$. After growth, a hexagonal lattice of
air holes was defined by electron beam lithography with a lattice constant of
$a=\unit{270}{\nano\meter}$ in a ZEP 520-A soft mask and deeply etched using a
SiCl$_\mathrm4$ based inductively coupled plasma to form a 2D PhC
\cite{Kress2005fabrication}. We incorporated an optimized L3 cavity design
\cite{yoshie2004vacuum,asano2006ultrahigh}, giving rise to cavities with
$V_{mode} \sim 0.92 (\nicefrac{\lambda}{n})^3$ and $Q= 8000 - 15000$. In a final
process step the AlGaAs layer was selectively removed with hydrofluoric acid to
establish a free standing membrane. 

For optical studies the sample was mounted
in a liquid He flow-cryostat and cooled to lattice temperatures of $T =
\unit{12}{\kelvin}$. Excitation of the sample was achieved via a $100\times$
high numerical aperture $NA=0.5$ confocal microscope objective that enables to
focus light to a diffraction limit spot with \mbox{$1/e^2$-size} of
$\unit{960}{\nano\meter}$. The optical response from the system was collected
via the same microscope objective and directly guided to a single imaging
monochromator and detected with a liquid nitrogen cooled CCD camera. For
measurements of the second order photon correlation function
\smash{$g^{(2)}(\tau)$}, the monochromator was used as a tunable bandpass filter
with a bandwidth of $\unit{270}{\micro\electronvolt}$. The spectrally filtered
signal was coupled into a fiber-beamsplitter and guided to two separate
avalanche photo diodes which provide single photon sensitivity and act as a
Hanbury Brown and Twiss setup \cite{brown1994correlation}. The avalanche photo
diodes exhibit a temporal resolution of $\unit{350}{\pico\second}$. The
detection events were time correlated using a Pico Quant TimeHarp time tagging
module. For mode resonant excitation
\cite{nomura2006localized,kaniber2009efficient} we used a continuous wave single
frequency laser with a bandwidth of $\unit{100}{\kilo\hertz}$ and a tuning range
between $\unit{1259}{\milli\electronvolt}$ and
$\unit{1369}{\milli\electronvolt}$.

\section{Auto- and cross-correlation spectroscopy}

\begin{figure}
    \includegraphics{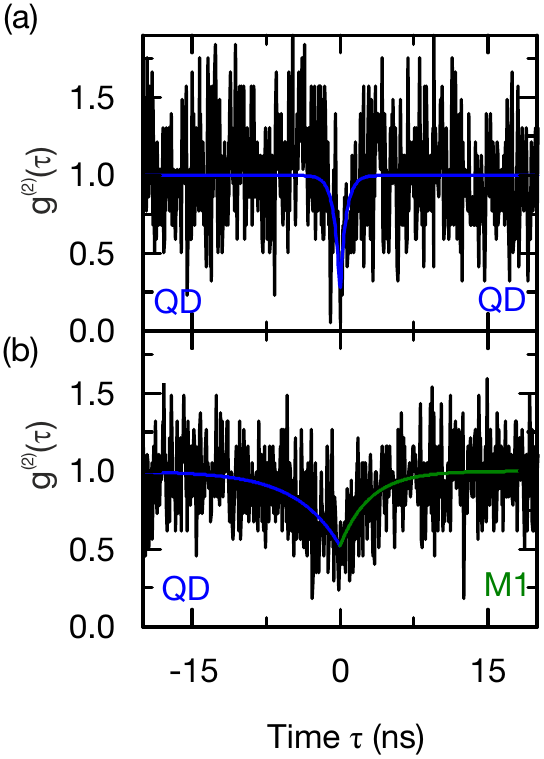}
    \renewcommand{\figurename}{Figure SM}
    \caption{\label{figure_SM1}%
    (a) Second-order autocorrelation measurement of QD emitting at $E_\mathrm{QD}=\unit{1263.1}{\milli\electronvolt}$ as shown in Figure 1a (main manuscript). (b) Second-order cross-correlation measurement between QD emitting at $E_\mathrm{QD}=\unit{1263.1}{\milli\electronvolt}$ and the cavitiy mode M1 emitting at $E_\mathrm{cav}=\unit{1257.1}{\milli\electronvolt}$ as shown in Figure 1a (main manuscript).}
\end{figure}
To gain more insight into the coupling between the spectrally detuned QD and the
cavity mode we performed measurements of the photon statistics of the emitted
light. First we measured the second order photon correlation function
$g^{(2)}(\tau)$ of the QD emitting at $\unit{1263.1}{\milli\electronvolt}$ with an
excitation power density close to saturation  ($\unit{\sim
0.14}{\kilo\watt\per\centi\meter\squared}$) as shown in figure SM\ref{figure_SM1} (a). For zero time delay $\tau=\unit{0}{\nano\second}$ we
observe a reduced number of correlation counts, giving rise to a value of
$g_{X-X}^{(2)}(0)=0.25 \pm 0.16$, demonstrating the non-classical character of the
studied quantum light and indicating that the signal stems pre-dominantly from a
single quantum emitter \cite{michler2000quantum,santori2001triggered}. As we
have proven the single photon characteristics of the emission we investigated
the coupling of the QD to the fundamental cavity mode M1 by performing cross-
correlation measurements between them. The measurement of the cross-correlation
function $g_{X-cav}^{(2)}(\tau)$ is presented in figure SM\ref{figure_SM1} (b)
for the same excitation power density of $\unit{\sim
0.14}{\kilo\watt\per\centi\meter\squared}$. As for the autocorrelation measurement of the QD, the
measurement between QD and cavity mode yields a strong suppression of
correlations for zero time delays with a $g_{X-cav}^{(2)}(0)=0.52 \pm 0.07$. This
antibunching unambiguously proves that the investigated QD is efficiently
coupled to the cavity mode, since we would expect an uncorrelated constant statistic
$g_{X-cav}^{(2)}(\tau)=1$ for an uncoupled QD and cavity
\cite{hennessy2007quantum,press2007photon,kaniber2008investigation}. This non-resonant cavity mode feeding has been shown to be due to a number of mechanisms
including coupling to acoustic phonons \cite{hohenester2009phonon} and higher
excited QD transitions \cite{laucht2010temporal}. The slightly increased value
of $g_{X-cav}^{(2)}(0)=0.52 \pm 0.07$ is attributed to additional sources of cavity
feeding, most probably due to other spectrally detuned QD transitions. The different
lifetimes $\tau_{auto}^{QD}= \unit{0.61 \pm 0.19}{\nano\second}$ and
$\tau_{cross}^{QD}=\unit{3.0 \pm 0.65}{\nano\second}$, extracted from the auto-correlation and cross-correlation measurement in figure SM\ref{figure_SM1}
(a) and (b), respectively, are due to different spectral cavity mode-QD
detunings $\Delta_{auto}^{cav- QD}=\unit{6.1}{\milli\electronvolt}$ and
$\Delta_{cros}^{cav-QD}= \unit{7.7}{\milli\electronvolt}$, respectively. We
conclude that we are investigating a state of the art nano-cavity doped with few
single photon emitters that are efficiently non-resonantly coupled to the cavity
mode via their multi-exciton states. Complete saturations of the cavity mode
emission proves that the cavity mode is only pumped by a few QD states.

\section{Theoretical Methods} 
\begin{figure}
    \includegraphics{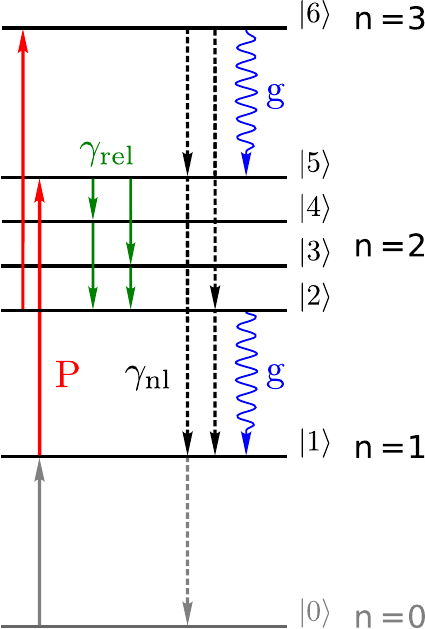}
    \renewcommand{\figurename}{Figure SM}
    \caption{\label{figure_SM2}%
    Schematic representation of the electronic QD level structure, corresponding to figure 1 (main manuscript) Light-matter coupling leads to recombination processes between the excitation manifolds with the light-matter coupling strength $g$. The arrows indicate the excitation,
    relaxation processses, and spontaneous emission into leaky modes taking place at a rate $P$, $\gamma_{\mathrm{rel}}$ and $\gamma_{\mathrm{nl}}$, respectively. Transitions between $n=1$ and $n=0$ are strongly off-resonant, thus the dynamics of the $\ket{0}$ state is effectively described by $\ket{1}$, see text.}
\end{figure}
In this section we discuss the theoretical model underlying the results shown in figure 2 (main manuscript) and provide details about the calculation of the $\beta(P)$-factor presented in figure 3 (main manuscript). For the theoretical description we consider an ensemble of $N_{\mathrm{QD}} = 4$ self-assembled QDs coupled to a single high-quality mode of the PhC cavity. 
Each QD $\alpha$ possesses a multitude of many-particle states $\ket{i_\alpha}$, represented by different excitation manifolds $n$, where $n$ stands for the total excitation number of e-h-pairs in the QD (compare figure 1 in the main manuscript). From these we choose six states ($\ket{1}\ldots\ket{6}$) that are numbered in order of increasing energy, as depicted in figure SM\ref{figure_SM2}. Optical transitions take place between states of manifolds $n$ and $n-1$ that differ by one e-h-pair, and we assume that two transitions ($\ket{6}\rightarrow \ket{5}$ and $\ket{2}\rightarrow \ket{1}$) are optically bright and resonant with the laser mode. 
A special role takes on the transition from the $n=1$ manifold to the ground state. In the experiment to which the model is applied, the ground-state exciton is strongly detuned from the cavity mode. In the configuration dynamics, the realization of the ground state is therefore strongly inhibited due to pumping, and the light-matter interaction is determined by the interplay of many-particle transitions between higher manifolds with $n>0$. Therefore, the $n=0$ state is not included in the calculations. 

The Hilbert space of the multi-QD and cavity-photon system is spanned by the product states $\ket{i_1}\cdots\ket{i_{N_{\mathrm{QD}}}}\ket{N}$, where $\ket{N}$ defines the $N$-photon Fock state of the cavity mode. Considering the QDs and the photon mode as an open quantum system, one can describe the dynamics of the system density operator by the von Neumann-Lindblad (vNL) equation
\begin{align}
    \frac{\partial}{\partial\hbox{t}} \rho &= -i[H_{\mathrm{JC}},\rho] +\mathcal{L}\rho .
    \label{eq:vNL}
\end{align}
The coherent dynamics is represented by the commutator with the Jaynes-Cummings (JC) Hamiltonian
\begin{align}
    H_{\mathrm{JC}} = g\sum_{\alpha,i}\left[b^\dagger D^{l}_{\alpha, i} + b\left(D^{l}_{\alpha, i}\right)^\dagger\right],
\end{align}
describing the non-perturbative light-matter interaction between the QD-interband transitions and the quantized field of the microcavity mode. Here, the operator
\begin{align}
    D^l_{\alpha, i} = \sum_{j} \ket{\alpha,j}\bra{\alpha,i}
\end{align}
expresses all dipole-allowed transitions of QD $\alpha$ between many-particle state $\ket{i_\alpha}$ and $\ket{j_\alpha}$ that are resonant with the laser mode, i.e. $\ket{6}\rightarrow \ket{5}$ and $\ket{2}\rightarrow \ket{1}$. $b^\dagger$ and $b$ are the bosonic creation and annihilation operators for photons in the laser mode, and $g$ is the light-matter coupling strength for the respective electronic states and the cavity mode. 

QDs are embedded systems, and their dynamics is strongly influenced by the coupling between localized QD states and the (quasi-) continuum of delocalized states and photon modes. While the former facilitates efficient carrier scattering into and within the QD via carrier–Coulomb and carrier–phonon interaction, the coupling to the latter leads to carrier recombination due to spontaneous emission. The incoherent, dissipative evolution results from the system-bath interaction, which leads to the sum of Lindblad operators
\begin{align}
    \mathcal{L}\rho &= \sum_X \frac{\gamma_X}{2}\left[ 2X\rho X^\dagger - X^\dagger X\rho - \rho X^\dagger X\right].
\end{align}
The operator $X$ describes a reservoir-assisted transition in the system. The relevant information about the bath and its interaction with the system is contained in the transition rate $\gamma_X$. In the system we consider, $X$ can be either a transition operator $\ket{i_\alpha}\bra{j_\alpha}$ between two QD many-particle states, or a photonic annihilation operator $b$. The first case corresponds to a change from many-particle state $\ket{j_\alpha}$ to $\ket{i_\alpha}$, including pump excitation, spontaneous recombination into non-lasing modes and relaxation processes at rates $P$, $\gamma_{\mathrm{nl}}$ and $\gamma_{\mathrm{rel}}$, respectively. All occurring incoherent processes are schematically depicted in figure SM\ref{figure_SM2}. In the second case the photon subsystem undergoes a lowering of the photon number due to cavity losses at a rate $\kappa$.

The von Neumann-Lindblad equation~\eqref{eq:vNL} is solved numerically until the steady-state solution is reached. Then various steady-state expectation values, such as the level populations and the photon statistics can be obtained. For the calculations shown in the main text, a light matter coupling strength of $g = 0.11$/ps is used, corresponding to a vacuum rabi splitting of $140\,\mu$eV. For the intraband relaxation rates $\gamma_{\mathrm{rel}}$ we use $0.5$/ps. Radiative losses into non-lasing modes are strongly suppressed in photonic crystal cavity devices, and we consider $\gamma_{\mathrm{nl}} = 0.01$/ps. The cavity decay rate of $\kappa = 0.16$/ps corresponds to a cavity-Q of $12000$ in the spectral range of the InGaAs QD emission. 

To calculate the pump-rate dependent spontaneous-emission coupling factor
\begin{align}
    \beta(P) = \frac{\Gamma(P)}{\Gamma(P) + \Gamma_{\mathrm{nl}}(P)}
	\label{eq:beta}
\end{align}
that is described by the ratio of the spontaneous emission into the lasing mode and the total spontaneous emission, at a given pump rate $P$ we consider the equation of motion for the cavity mean photon number 
\begin{align}
    \frac{d}{d\hbox{t}} \braket{b^\dagger b} = - \kappa \braket{b^\dagger b} + \Gamma + \Gamma_{\mathrm{stim}}.
    \label{eq:eom_bTb}
\end{align}
It is determined by the balance between cavity losses in the first term and photon emission. The second term constitutes the total spontaneous emission into the laser mode
\begin{align}
    \Gamma = \Gamma_{\mathrm{spont}} + \Gamma_{\mathrm{sr}}.
\end{align}
Here $\Gamma_{\mathrm{spont}}$ is the usual contribution from independent emitters to the spontaneous emission, and $\Gamma_{\mathrm{sr}}$ reflects the enhancement or suppression of spontaneous emission due to QD-QD correlations.
The third term in Eq.~\eqref{eq:eom_bTb} represents the contribution due to stimulated emission and absorption.

In the steady state we obtain 
\begin{align}
    \Gamma_{\mathrm{spont}} &= \sum_{\alpha=1}^{N_{\mathrm{QD}}} \sum_i R_i \Braket{\left(D^{l}_{\alpha,i}\right)^\dagger D^{l}_{\alpha,i}}
\end{align}
and
\begin{align}
    \Gamma_{\mathrm{sr}} &= \sum_{\alpha\neq\beta}^{N_{\mathrm{QD}}} \sum_{i,j} R_i \Braket{D^{l}_{\alpha,i}\left(D^{l}_{\beta,j}\right)^\dagger}.
\end{align}
For more details we refer to Ref.~\cite{leymann2015sub}. The quantity $R_i$ is the spontaneous emission rate for recombination of the many-particle state $\ket{i_\alpha}$
\begin{align}
    R_i = \frac{4g^2}{\kappa + \gamma^{\mathrm{tot}}_{i}},
\end{align}
which resembles the form known from rate equations, where $\gamma^{\mathrm{tot}}_{i}$ is the total dephasing of the many-particle transition with the initial state $\ket{i_\alpha}$. Similarly, an expression for the emission into non-lasing modes can be found
\begin{align}
    \Gamma_{\mathrm{nl}} = \gamma_{\mathrm{nl}} \sum_{\alpha=1}^{N_{\mathrm{QD}}} \sum_i \Braket{\left(D^{\mathrm{nl}}_{\alpha,i}\right)^\dagger D^{\mathrm{nl}}_{\alpha,i}}.
    \label{eq:gamma_nl}
\end{align}
In this expression $D^{\mathrm{nl}}_{\alpha,i}$ describes all dipole-allowed transitions of QD $\alpha$ that are detuned from the laser mode.
Eqns.~\eqref{eq:beta} - \eqref{eq:gamma_nl} are explicitly evaluated to obtain the pump-power dependent $\beta(P)$-factor introduced in the main text. All averages are taken with respect to the steady state density matrix.

\FloatBarrier
\bibliographystyle{apsrev}
\bibliography{Bibliography_BunchingCavity}